\documentclass[prb,amsmath,amssymb,twocolumn]{revtex4}
\usepackage[dvips]{graphicx}
\usepackage{dcolumn}
\usepackage{bm}

\def\dd{\textrm{d}}

\begin{document}
\preprint{cond-mat/0609106}
\title{Renormalization group fixed points, universal phase diagram,
and $1/N$ expansion for\\ quantum liquids with interactions near the
unitarity limit}

\author{Predrag Nikoli\'c}
\affiliation{Department of Physics, Harvard University, Cambridge MA
02138}

\author{Subir Sachdev}
\affiliation{Department of Physics, Harvard University, Cambridge MA
02138}

\begin{abstract}
It has long been known that particles with short-range {\em
repulsive\/} interactions in spatial dimension $d=1$ form universal
quantum liquids in the low density limit: all properties can be
related to those of the spinless free Fermi gas. Previous
renormalization group (RG) analyses demonstrated that this
universality is described by an RG fixed point, infrared stable for
$d<2$, of the zero density gas. We show that for $d>2$ the {\em
same\/} fixed point describes the universal properties of particles
with short-range {\em attractive\/} interactions near a Feshbach
resonance; the fixed point is now infrared unstable, and the
relevant perturbation is the detuning of the resonance. Some
exponents are determined exactly, and the same expansion in powers
of $(d-2)$ applies for scaling functions for $d<2$ and $d>2$. A
separate exact RG analysis of a field theory of the particles
coupled to `molecules' finds an alternative description of the same
fixed point, with identical exponents; this approach yields a
$(4-d)$ expansion which agrees with the recent results of Nishida
and Son (Phys. Rev. Lett. {\bf 97}, 050403 (2006)). The existence of
the RG fixed point implies a universal phase diagram as a function
of density, temperature, population imbalance, and detuning; in
particular, this applies to the BEC-BCS crossover of fermions with
$s$-wave pairing. Our results open the way towards computation of
these universal properties using the standard field-theoretic
techniques of critical phenomena, along with a systematic analysis
of corrections to universality. We also propose a $1/N$ expansion
(based upon models with Sp($2N$) symmetry) of the fixed point and
its vicinity, and use it to obtain results for the phase diagram.
\end{abstract}

\date{\today}

\maketitle

\section{Introduction}
\label{sec:intro}

The study of ultracold atomic gases has drawn renewed attention to
interacting quantum gases in regimes where their properties are
independent of all microscopic details of the interaction between
the atoms. This happens when the scattering cross-section between
two atoms reaches its unitarity limit.

The simplest example of this phenomenon is the dilute Bose gas with
short-range repulsive interactions in spatial dimension $d=1$ where,
in fact, the properties are generically universal. It is a
well-known property of quantum-mechanical scattering in one
dimension that for any generic short-range potential, the $S$ matrix
for two particle scattering reaches its unitarity limit value of
$-1$ in the limit of zero momentum transfer. However, the strongly
interacting Bose gas with $S=-1$ can be reinterpreted \cite{tonks}
as a {\em free\/} Fermi gas (the `Tonks' gas), allowing easy
computation of at least the thermodynamic properties. Correlation
functions are much harder to compute, especially at finite
temperature and unequal times, but much theoretical effort has been
expended in this direction.\cite{korepin,kedar,apy,book}
Experimental studies of Tonks gas behavior have also
appeared.\cite{bloch}

No generic unitarity limit scattering is obtained in $d=3$. In this
case, we usually have the limit $S=1$ for low momentum, implying
that sufficiently slowly moving particles do not scatter. However,
upon fine-tuning the strength of an attractive interaction, it is
possible to obtain unitarity limit scattering across a Feshbach
resonance.\cite{bertsch,ho,jin} Near a Feshbach resonance, the
$s$-wave scattering amplitude, $f_0$, takes the form\cite{rembert}
\begin{equation}
f_0 = \frac{1}{\nu - ik} \label{fr1}
\end{equation} where $k$ is the momentum transfer,
$S=1+2ik f_0$, and $\nu$ is the `detuning' across the resonance (it
is related to the scattering length, $a$, by $\nu = -1/a$). The
resonance occurs at $\nu=0$, and notice then that $S=-1$, the
unitarity limit. We will be interested here in {\em broad\/}
Feshbach resonances, where Eq.~(\ref{fr1}) holds over all the
momenta of interest in a quantum gas. For a gas with density $n$, a
resonance is broad if the $\mathcal{O}(k^2)$ corrections to the
denominator of Eq.~(\ref{fr1}) are negligible at the characteristic
momentum $n^{1/3}$ (we use $\hbar=1$ throughout). Notice then that
the width condition on the Feshbach resonance, is actually a
diluteness condition on the quantum gas, and any Feshbach resonance
is `broad' for a sufficiently dilute gas.

This paper demonstrates that a unified understanding of the distinct
universal properties of the quantum gases in $d=1$ and $d=3$ can be
obtained in a field-theoretic renormalization group (RG) analysis.
The RG approach was used previously\cite{kolo,sss,kedar,book} for
the repulsive Bose gas in $d<2$, and here we demonstrate that a
direct generalization applies to a wide class of quantum liquids in
general $d$. We will use the structure of an RG fixed point to argue
that there is a {\em universal phase diagram\/} as a function of
relevant perturbations away from the fixed point: these
perturbations are the temperature, the detuning away from the
Feshbach resonance, and the density imbalance for a two-component
quantum gas. Our claims of universality are stronger than previous
ones which applied only at the resonance.\cite{bertsch,ho,jin} In
particular, questions on the nature of the phases, including the
existence of exotic gapless superfluid states, have a unique and
universal answer for a broad Feshbach resonance. We will also obtain
explicit results for this phase diagram, and for associated
universal scaling functions, for the two-component Fermi gas in
$d=3$ with $s$-wave pairing using a $1/N$ expansion of a model with
Sp($2N$) symmetry. In particular, we obtain results on the equation
of state for the normal state with unbalanced densities and two
Fermi surfaces.

The form of our universal phase diagram has similarities to work by
Sheehy and Radzihovsky\cite{leo0,leo} (and related work by
others\cite{yip}) where they departed from the narrow resonance
limit in an expansion in the resonance width. We discover the same
phases and transitions between them, except the possible non-uniform
phases which are not a focus of this paper. The main advance in our
work is a systematic treatment of the broad resonance limit, which
is experimentally relevant, and in which universality applies.
Universality has also been explored using a different RG analysis in
Ref.~\onlinecite{diehl}: they do not give a ``critical phenomena''
interpretation of the results, which we claim below yield the most
direct interpretation and computation tools for identifying
universal behavior.

The previous study\cite{kolo,sss,kedar,book} of the dilute Bose gas
began with an RG analysis of the field theory of the zero density
gas\cite{fwgf} at chemical potential $\mu=0$, considered as a
quantum critical point between the vacuum state for $\mu < 0$, and
the finite density ground state for $\mu > 0$. Interactions between
a pair of particles created out of the vacuum are characterized by a
coupling $u$ (defined more precisely in the body of the paper), and
this coupling was shown to obey the {\em exact\/} field-theoretic RG
equation
\begin{equation}
\frac{du}{d\ell} = (2-d) u - \frac{u^2}{2} \label{rg1}
\end{equation}
under a rescaling of length scales by a factor of $e^\ell$, and a
dynamic exponent $z=2$. As we will review later, this RG equation
applies equally to two-body interactions between particles with
arbitrary statistics and masses.
\begin{figure}[tb]
  \includegraphics[width=3in]{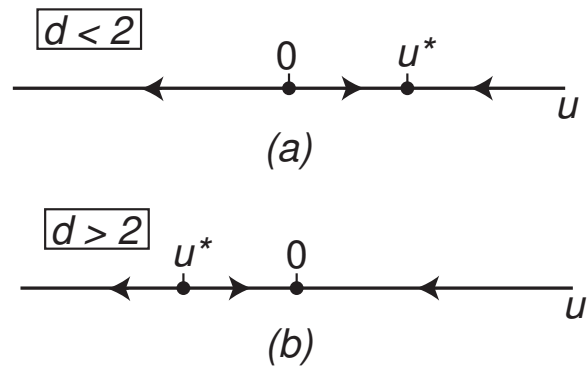}
  \caption{The exact RG flow of Eq.~(\ref{rg1}), as discussed in
  Ref.~\onlinecite{book}. Here $u$ is a measure of the short-range
  two-body interaction between the particles, and the RG applies in
  the limit of low density of either Bose, Fermi, or Bose-Fermi
  quantum liquids.
  ({\em a\/}) For $d<2$, the infrared stable fixed point
  at $u=u^\ast > 0$ describes quantum liquids of either bosons or fermions
  with repulsive interactions which are
  generically universal in the low density limit. In $d=1$ this fixed point is described
  by the spinless free Fermi gas (`Tonks' gas), for all statistics and spin of the
  constituent particles. ({\em b\/}) For $d>2$, the
  infrared unstable fixed point at $u=u^\ast < 0$ describes the Feshbach
  resonance which obtains for the case of attractive interactions. The relevant perturbation
  $(u-u^\ast)$ corresponds to the the detuning from the resonant
  interaction.
  }
\label{rgflow}
\end{figure}
As shown in Fig.~\ref{rgflow}, for $d<2$, this RG equation has an
infrared stable fixed point at $u^\ast >0$. It was argued
\cite{kolo,book,sss} that this fixed point interaction which
describes unitarity scattering. The chemical potential, $\mu$, is a
relevant perturbation at this fixed point, and its scaling
dimension, and RG flow can also be determined exactly. A standard
\cite{brezin} field-theoretic analysis then allowed computation
\cite{book} of all universal properties of the $\mu \neq 0$ Bose gas
in $d<2$, in an expansion which can be carried out to all orders in
$u^\ast = 2(2-d)$. The degree of difficulty of this computation is
the same for thermodynamic and dynamic correlations. Practically
speaking, this expansion is then useful for those physical
properties which are difficult to access by the fermionization
method in $d=1$. In some cases, an exact computation is possible
both in the fermionization approach, and also by resumming all
orders in the $(2-d)$ expansion, and exact agreement was
obtained.\cite{kedar} A review of this approach to Bose and Fermi
quantum liquids in $d<2$ appears in Section~\ref{sec:a}.

Let us now examine the RG equation (\ref{rg1}) for $d>2$. The flows
are shown in Fig~\ref{rgflow}. Now there is an infrared {\em
unstable\/} fixed point at a $u^\ast < 0$. We show here that this
fixed point describes the Feshbach resonance of particles with
attractive interactions. The unitarity of the scattering at this
fixed point has also been pointed out independently in
Ref.~\onlinecite{kopi}. The relevant perturbation $(u-u^\ast)$ is
proportional to the detuning, $\nu$, of the resonance. By an
elementary computation of the eigenvalue of linear perturbations
from the $u=u^\ast$ fixed point of Eq.~(\ref{rg1}) we therefore
determine the exact scaling dimension
\begin{equation}
\mbox{dim}[\nu] = d-2
\end{equation}
The universal properties of the quantum liquid with $\nu \neq 0$ and
$\mu \neq 0$ can now be determined by essentially the same
renormalized theory earlier used for $d<2$, and now becomes an
expansion in $(d-2)$. Indeed, as we will see in Section~\ref{sec:a},
the earlier\cite{book} $(2-d)$ expansions for scaling functions for
$d<2$, apply {\em unchanged\/} at resonance for $d>2$. A shortcoming
of this $(d-2)$ expansion for the attractive Fermi gas is that the
pairing amplitude is of order $\exp (1/u^\ast ) \sim \exp
(-1/(d-2))$, and so all effects associated with superconductivity
are non-perturbative.

\begin{figure}
\includegraphics[width=3in]{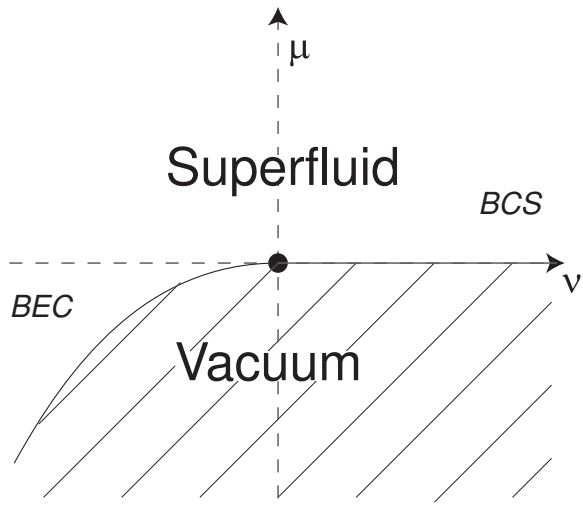}
\caption{\label{zeroh} Universal phase diagram at zero temperature
($T=0$) and balanced densities ($h=0$) for the two-component Fermi
gas in $d=3$. The vacuum state (shown hatched) has no particles, and
is present for $\mu<0$ and $\nu>0$, or for $\nu<0$ and $\mu < -\nu^2
/(2m)$, where $m$ is the mass of a fermion. The position of the
$\nu<0$ phase boundary is determined by the energy of the
two-fermion bound state. The density of particles vanishes
continuously at the second order quantum phase transition boundary
of the superfluid phase, which is indicated by the thin continuous
line. The quantum multicritical point at $\mu=\nu=h=T=0$ (denoted by
the filled circle) is the RG fixed point which is the basis for the
analysis in this paper.}
\end{figure}

Regardless of the ability to determine scaling functions in the
$(d-2)$ expansion, the flow equation (\ref{rg1}), and its associated
exponents are exact, and allow us to deduce some exact scaling forms
associated with the RG fixed point in $d>2$. For the two-component
Fermi gas, we have already noted two relevant perturbations at the
fixed point, the detuning $\nu$ and the chemical potential $\mu$.
For general $d$, we define the overall scale of $\nu$ by the
requirement that the $s$-wave scattering amplitude is proportional
to $1/(\nu - (-k^2)^{(d-2)/2})$. The scaling dimension of $\mu$ is
exactly 2. There is also a third relevant perturbation at the fixed
point: a `field' $h$ conjugate to the density difference of the two
species of fermions, which also has scaling dimension 2. From all
this information, and the dynamic exponent $z=2$, we can deduce one
of our central results, a scaling form for the grand canonical free
energy density $\mathcal{F}$ (from the usual thermodynamic identity,
$\mathcal{F}=-P$, the pressure) at a temperature $T$:
\begin{equation}
\mathcal{F} = (2m)^{d/2} T^{1+d/2} F_\mathcal{F} \left(
\frac{\mu}{T}, \frac{h}{T}, \frac{\nu}{(2mT)^{(d-2)/2}} \right),
\label{scalef}
\end{equation}
where $m$ is the mass of one of the particles, and $F_\mathcal{F}$
is a universal scaling function whose three arguments extend over
all real values (if the mass of the particles are distinct, the
scaling function will also depend upon the ratio of the masses). For
the two-component Fermi gas this universal scaling function contains
all the information on the BEC-BCS
crossover.\cite{nsr,mohit,levin,griffin} The existence of this
scaling form also implies that there is a universal phase diagram as
a function of $\mu$, $\nu$, $h$, and $T$ in the low density limit
(or equivalently, for a broad resonance). Information on all the
phases and phase transitions in this phase diagram is also contained
in Eq.~(\ref{scalef}). At $h=0$ and $T=0$, the form of this
universal phase diagram can be deduced exactly, and is shown in
Fig.~\ref{zeroh} for $d=3$.
There is a second-order quantum phase transition line separating the
zero density vacuum from the low density superfluid. The RG fixed
point is a multicritical point on this line, and the whole phase
diagram is constructed as a theory of relevant perturbations from
this point. Note that there is no phase transition at finite density
within the superfluid, only a smooth crossover between the BCS and
BEC limits. However, there is a quantum critical point at zero
detuning (at ``resonance'') on the phase boundary, and we use this
critical point as a point of departure to describe the entire phase
diagram. We will present further results on this phase diagram in
the body of the paper.

Given the limitations of the $(d-2)$ expansion for obtaining useful
approximations for the scaling function, the remainder of the paper
presents alternative approaches to analyzing the fixed point
described above. In Section~\ref{sec:am}, we examine another field
theory commonly used to describe a Feshbach resonance: a theory of
mixed `atoms' and `molecules', with the `molecule' corresponding to
the bound state in the vicinity of the Feshbach resonance (the
so-called ``two-channel'' approach \cite{twochannel,leo}). As argued
above, it is useful to perform an RG analysis in the limit of zero
density, when the physics reduces to the quantum mechanics of
few-particle scattering. The RG flow equations can now also be
determined exactly, and are presented in Section~\ref{sec:am}. They
contain a fixed point, which, at first glance, appears distinct from
that of Eq.~(\ref{rg1}). However, a careful analysis of its
universal properties, such as scaling dimensions of operators, shows
that they are identical to the results quoted above. The finite
density field theory associated with this fixed point becomes weakly
coupled when $(4-d)$ is small. Consequently, a standard renormalized
perturbative expansion now allows us to obtain universal properties
in powers of $(4-d)$. The universal results of the $(4-d)$ expansion
are equal to those obtained recently in the important work of
Nishida and Son \cite{son}, who motivated the expansion from
different considerations and use a different computational scheme.
The special role of $d=2$ and $d=4$ was also discussed by Nussinov
and Nussinov.\cite{nussinov} Our method streamlines the expansion
$(4-d)$ onto the standard approach used extensively for critical
phenomena, and also opens the way to a systematic analysis of
corrections to scaling by a consideration of irrelevant
perturbations.

Further considerations on the structure of the $(4-d)$ expansion
shows that it also has some limitations in its range of validity.
For the two component Fermi gas, the naive $(4-d)$ expansion is
restricted to being well within the superfluid state, or well within
the normal state with one Fermi surface. A subtle redefinition of
parameter scales \cite{son,son2} is required to obtain an expansion
for the 2 Fermi surface normal state. The latter is one of the most
interesting new regimes discovered in recent
experiments\cite{martin,randy}, and so an unrestricted analysis here
would be useful.

With this motivation, in Section~\ref{sec:largeN} we introduce a
{\em third\/} approach towards analyzing the finite density field
theory at the RG fixed point of Eq.~\ref{rg1}. This is based upon
the $1/N$ expansion, where $N$ is the number of particle species.
This expansion holds uniformly for general $\mu$, $\nu$, $h$, and
$T$. For the two-component Fermi gas, the critical fixed point has
Sp($2N$) symmetry, and the expansion is based upon that considered
earlier in Ref.~\onlinecite{wang} well away from the Feshbach
resonance.

We conclude this introduction by briefly discussing quantum liquids
with attractive interactions in $d<2$; these are not directly
described by the fixed points discussed so far. In such systems, the
`atomic' constituents will bind to form `molecules' and the
resulting molecules could have repulsive interactions, leading
eventually to the quantum liquid of the type discussed above for
$d<2$. The crossovers associated with this behavior will be
described by an RG analysis which is a combination of that presented
in Sections~\ref{sec:a} and~\ref{sec:am}---we need to explicitly
include the interactions between the atoms and the molecules
(relevant for $d<2$), along with 3-point term allowing for a
molecule to decay into two atoms (relevant for $d<4$). We will not
present the details of such an RG analysis here, but note that these
considerations are entirely consistent with a recent analysis by
Gurarie \cite{gurarie} in $d=1$ using Bethe ansatz methods.

\section{RG for a field theory of atoms}
\label{sec:a}

We begin by considering the theory of the dilute Bose gas, from the
perspective of a zero density quantum critical point, as discussed
in Ref.~\onlinecite{book}. We consider a zero density theory at the
chemical potential $\mu=0$. This can be viewed as a quantum critical
point between the finite density phase with $\mu > 0$, and the empty
vacuum ground state for $\mu < 0$. The critical theory of this
quantum critical point of a boson, $\psi$, is
\begin{eqnarray}
\mathcal{S}_{c}^{b} = \int \dd \tau \dd^d x \Biggl\{ \psi^\dagger \left(
\frac{\partial}{\partial \tau} - \frac{\nabla^2}{2 m} \right)\psi +
\frac{u_0}{2} \psi^\dagger \psi^\dagger \psi \psi \Biggr\}.
\end{eqnarray}

The RG analysis of the zero density theory does not depend upon the
statistics of the particles, so we also consider in parallel a
theory of two-component fermions, $\psi_\sigma$, $\sigma=1,2$. The
critical theory is
\begin{eqnarray}
\mathcal{S}_{c}^{f} = \int \dd \tau \dd^d x \Biggl\{ \psi^\dagger_\sigma
\left( \frac{\partial}{\partial \tau} - \frac{\nabla^2}{2 m}
\right)\psi_\sigma + u_0 \psi^\dagger_1 \psi^\dagger_2 \psi_2 \psi_1
\Biggr\}.
\end{eqnarray}

The single particle self-energy is identically zero at $T=0$ because
the vacuum state contains no particles. Consequently, we can perform
an RG transformation with the exact dynamic critical exponent $z=2$,
and obtain the exact scaling dimension
\begin{equation}
\mbox{dim}[\psi] = \mbox{dim}[\psi_\sigma] = d/2. \label{rg2}
\end{equation}

Despite the trivial nature of the vacuum at $T=0$, there is a
non-trivial RG transformation of the two body interaction $u_0$. This
comes entirely from the ladder diagram of repeated two-particle
scattering. Defining a renormalized dimensionless coupling $u$ by
\begin{equation}
u_0 = \frac{\kappa^{2-d}Z_4}{2m S_d} u,
\end{equation}
where $\kappa$ is a renormalization momentum scale, $S_d =
2/(\Gamma(d/2) (4 \pi)^{d/2})$ is the standard phase space factor,
and $Z_4$ is the renormalization constant. The ladder diagrams lead
to the exact result
\begin{equation}
Z_4^{-1} = 1-\frac{u}{2-d}
\end{equation}
in the minimal subtraction scheme. From this, we obtain the exact RG
$\beta$ function for $u$ presented earlier in Eq.~(\ref{rg1}), upon
rescaling $\kappa$ by a factor of $e^\ell$.

Now we include relevant operators away from this fixed point (apart
from $(u-u^\ast)$ for $d>2$). For the single-component Bose gas,
there is only one such term, the chemical potential
\begin{eqnarray}
\mathcal{S}_{bp} = \int \dd \tau \dd^d x \Biggl\{ - \mu \psi^\dagger
\psi \Biggr\}.
\end{eqnarray}
It easy to see that an insertion of $\mathcal{S}_p$ acquires no
operator renormalization at the zero density critical point, and so
from Eq.~(\ref{rg2}) we conclude that
\begin{equation}
\mbox{dim}[\mu] = 2.
\end{equation}
For the two-component Fermi gas, we have an additional relevant
perturbation corresponding to the difference in chemical potential
of the two species:
\begin{eqnarray}
\mathcal{S}_{fp} = \int \dd \tau \dd^d x \Biggl\{ &-& \mu
(\psi^\dagger_1 \psi_1 + \psi_2^\dagger \psi_2) \nonumber \\ &-& h
(\psi^\dagger_1 \psi_1 - \psi_2^\dagger \psi_2)  \Biggr\},
\end{eqnarray}
and again we have
\begin{equation}
\mbox{dim}[h]=2
\end{equation}

Now let us examine the universal properties of the finite density
$\mu \neq 0$ theory in the vicinity of the fixed point of
Eq.~(\ref{rg1}). We will consider the cases $d<2$ and $d>2$
separately in the following subsections.

\subsection{$d<2$}
\label{dl2}

The fixed point of Eq.~(\ref{rg1}) has repulsive interactions with
$u^\ast = 0$.

The properties of the Bose gas at this fixed point were already
considered in Refs.~\onlinecite{kolo,book}. As an example, we note
the grand canonical free energy density at $T=0$, {\em i.e.\/} the
pressure $P$. From Eq.~(\ref{scalef}), we deduce that this obeys
\begin{equation}
P = C_{d}^b \mu (2 m \mu)^{d/2} \label{p1}
\end{equation}
where $C_{d}^b$ is a universal number. A $(2-d)$ expansion for
$C_d^b$ was presented in Ref.~\onlinecite{book}, and the leading
result is \cite{book}
\begin{equation}
C_{d}^b = S_d \left( \frac{1}{4 (2-d)} + \frac{2 \ln 2 -1}{16} +
\mathcal{O} (2-d) \right).
\end{equation}

The corresponding universal properties of the two-component Fermi
gas at the repulsive $u=u^\ast$ are also easy to work out. For
simplicity, we present results only at $T=0$ and $h=0$. Then, the
result (\ref{p1}) still applies, but with a different universal
constant $C_d^f$. This is easily computed from the perturbation
theory for the pressure of the two-component Fermi gas to the first
order in $u_0$:
\begin{eqnarray}
P = && 2 \int_0^{\sqrt{2m\mu}} \frac{\dd^d k}{(2 \pi)^d} \left(
\frac{k^2}{2m} - \mu \right) \nonumber \\ &&+ u_0 \left[
\int_0^{\sqrt{2m\mu}} \frac{\dd^d k}{(2 \pi)^d}  \right]^2 +
\mathcal{O}(u_0^2) \label{p2}
\end{eqnarray}
The universal result is obtained by evaluating (\ref{p2}) at
$u=u^\ast$ consistently to order $(2-d)$.
\begin{equation}
C_{d}^f = S_d \left( \frac{1}{2}  -  \frac{(2-d)}{8}  + \mathcal{O}
(2-d)^2 \right). \label{cdf}
\end{equation}

Of course, in both the Bose and Fermi cases considered in this
subsection, we can obtain exact results in $d=1$. The thermodynamics
of both models are equivalent to the free, spinless, Fermi gas. Note
that even the thermodynamics of the dilute two-component Fermi gas
is given by the spinless Fermi gas. From this, we easily obtain
\begin{equation}
C_1^b = C_1^f = \frac{2}{3 \pi}
\end{equation}

\subsection{$d>2$}
\label{dg2}

The fixed point of Eq.~(\ref{rg1}) is now at $u^\ast = - 2(d-2) <
0$.

The Bose gas with such attractive interactions is unstable, and
further interactions are needed to stabilize the theory: we will
therefore not consider it further here.

The Fermi gas is stable and has universal properties that can be
computed for small $(d-2)$. Indeed, the expansion described in
Section~\ref{dl2} applies {\em unchanged\/} also for $d>2$. In
particular, the pressure is still given by Eq.~(\ref{cdf}), and this
agrees with a recent result \cite{son2} obtained while this paper
was in preparation. Of course, as noted in Section~\ref{sec:intro},
this expansion knows nothing about pairing between the fermions, as
such effects are \cite{son2} exponentially small in $1/(2-d)$.

\section{RG for a field theory of atoms and molecules}
\label{sec:am}

This section will use the popular ``two-channel'' formulation of the
Feshbach resonance \cite{twochannel,leo} to obtain an alternative
field theoretic description of universal fixed point of
Eq.~(\ref{rg1}). In addition to the `elementary' particles,
$\psi_{1,2}$, we will also allow for a `composite' molecule, $\Phi$.
For generality, we will now allow the masses of the $\psi_\sigma$
particles to be possibly distinct, $m_1$ and $m_2$. By Galilean
invariance, the mass of the `composite' particle is $m_1 + m_2$. The
statistics of the two $\psi_\sigma$ can be arbitrary---they can be
be Fermi-Fermi, Bose-Fermi\cite{powell}, or Bose-Bose\cite{stoof}.
The `composite' particle $\Phi$ is then, naturally, a boson,
fermion, or a boson respectively.

Our analysis will find a `critical' fixed point describing the
unitarity limit theory. All exponents and scaling dimensions
associated with the fixed point will turn out to be identical to
those obtained for a seemingly distinct fixed point in
Section~\ref{sec:a}, suggesting that the two theories are in fact
identical.

As in Section~\ref{sec:a}, it is useful to begin with an RG analysis
of the zero density `critical' theory, which is presented in
Section~\ref{sec:zero}. The perturbations away from the critical
point are considered in Section~\ref{sec:perturb}.

\subsection{Zero density critical theory}
\label{sec:zero}

The action of the critical theory is now
\begin{eqnarray}
\mathcal{S}_c = && \int \dd \tau \dd^d x \Biggl\{ \psi_\sigma^\dagger
\left( \frac{\partial}{\partial \tau} - \frac{\nabla^2}{2 m_\sigma}
\right) \psi_\sigma \nonumber \\
&&~~~~~~~~+ \Phi^\dagger \left( \frac{\partial}{\partial \tau} -
\frac{\nabla^2}{2 (m_1 + m_2) } + r_{0c} \right) \Phi \nonumber \\
&&~~~~~~~~- g_0 \left( \Phi^\dagger \psi_1 \psi_2 + \psi_2^\dagger
\psi_1^\dagger \Phi \right) + \ldots \Biggr\}
\end{eqnarray}
Here $r_{0c}$ is a cutoff-dependent bare ``mass'' term which has to be
tuned to place the theory at its critical point {\em i.e.\/}
describe the unitarity limit scattering. In dimensional
regularization, $r_{0c}=0$. A tree-level scaling analysis with $z=2$
shows that
\begin{equation}
\mbox{dim}[\psi_\sigma] = d/2~~;~~\mbox{dim}[\Phi] =
d/2~~;~~\mbox{dim}[g_0] = (4-d)/2
\end{equation}
All other interactions are irrelevant for $d>2$. The scaling
dimension of $g_0$ naturally suggests that we perform a
renormalization group analysis as an expansion in $(4-d)$.

It turns out that it is possible to perform an RG analysis {\em
exactly\/}, to all orders in $(4-d)$. Most of the Feynman diagrams
vanish because of the zero density of the particles. The only
non-zero renormalization is a wavefunction renormalization of the
composite particle $\Phi$. The wavefunction renormalization is
computed from the $\Phi$ Green's function
\begin{equation}
G_\Phi (k, \omega) = \frac{1}{-i \omega + k^2 /(2 (m_1 + m_2 )) +
r_{0c} - \Sigma_\Phi (k, \omega)}
\end{equation}
For the above theory, at $T=0$, the exact expression for the self
energy is
\begin{eqnarray}
\Sigma _\Phi (k, \omega) &=&  g_0^2 \int \frac{ \dd \Omega}{2 \pi}
\int \frac{\dd^d p}{(2 \pi)^d} \frac{1}{(- i \Omega + p^2 /(2 m_2))}
\nonumber \\ &~&~~~~~~~~\times \frac{1}{ ( - i(\omega - \Omega) +
(k-p)^2 /(2 m_1))}
\nonumber \\
&=&  g_0^2 \frac{ \Gamma (1-d/2)}{(4 \pi)^{d/2}} \left[ \frac{2 m_1
m_2}{(m_1 + m_2 )} \right]^{d/2} \nonumber \\
&~&~~~~~~\times \left[ -i \omega + \frac{k^2}{2 (m_1 + m_2)}
\right]^{d/2-1}. \label{gp}
\end{eqnarray}
In the last equation, we have dropped a constant cutoff dependent
term which cancels against $r_{0c}$ (additional subtractions are
necessary for $d>4$). We now define a renormalized coupling $g$ by
\begin{equation}
g_0 =  \kappa^{(4-d)/2} \left[ \frac{2 m_1 m_2}{(m_1 + m_2 )}
\right]^{-d/4}\frac{1}{\sqrt{Z_\Phi S_d}}g. \label{defg}
\end{equation}
where $Z_\Phi$ allows for a non-trivial wavefunction renormalization
of the $\Phi$ field. It is easy to check that additional self-energy
and vertex corrections vanish in the zero density theory at $T=0$,
and so there is no wavefunction renormalization for the
$\psi_\sigma$, nor an independent coupling constant renormalization
of $g_0$.  In terms of $g$, we can expand the self energy to leading
order in $(4-d)$, and obtain
\begin{equation}
\Sigma_\Phi (k, \omega) = - \frac{g^2}{Z_\Phi (4-d)} \left[ -i
\omega + \frac{k^2}{2 (m_1 + m_2)} \right]
\end{equation}
Cancelling poles in the minimal subtraction scheme\cite{PeskinSchroeder},
we obtain the exact value of $Z_\Phi$:
\begin{equation}
Z_\Phi = 1 - \frac{g^2}{(4-d)} \label{valZ}
\end{equation}
Using Eq.~(\ref{defg}), we can compute the exact $\beta$ function
for $g$:
\begin{equation}
 \frac{dg}{d\ell} =  \frac{(4-d)}{2} g - \frac{g^3}{2}. \label{rgg}
\end{equation}
Also, from Eq.~(\ref{valZ}), we obtain for the anomalous dimension
of the composite field $\Phi$:
\begin{equation}
\eta (g) = g^2.
\end{equation}

For $d<4$, the RG flow equation Eq.~(\ref{rgg}) has an infrared {\em
stable\/} fixed point at $g^\ast = \sqrt{(4-d)}$. At first sight,
this may seem puzzling because in Section~\ref{sec:a} we described
the Feshbach resonance by an infrared {\em unstable\/} fixed point.
However, this is just a matter of our restriction here to a
particular critical manifold. We will see in
Section~\ref{sec:perturb}, where we consider the full parameter
space of the theory, that there is indeed an additional relevant
operator here which corresponds to the relevant perturbation of
Section~\ref{sec:a}, and this is the detuning away from the Feshbach
resonance. The attractive flow towards the $g=g^\ast$ fixed point
above corresponds to the $\mathcal{O}(k^2)$ corrections to the
scattering amplitude we discussed below Eq.~(\ref{fr1}): the system
can be assumed to be exactly at $g=g^\ast$ for a broad Feshbach
resonance, while the irrelevant flows towards $g=g^\ast$ have to be
included for a narrow resonance. As stated earlier, we restrict our
computations here to the broad resonance.

The critical exponents are to be evaluated at the fixed point of
Eq.~(\ref{rgg}). From this, we obtain the exact results
\begin{equation}
\eta = \eta (g^\ast ) = (4-d)~~;~~\mbox{dim}[\Phi] = (d+\eta)/2 = 2.
\label{exp}
\end{equation}
This (along with Galilean invariance) implies that the $\Phi$
Green's function behaves like $G_\Phi^{-1} \sim [-i \omega + k^2/(2
(m_1+m_2))]^{1-\eta/2}$, which is, of course, the result already
obtained in Eq.~(\ref{gp}).

At this point, the above RG fixed point appears to bear little
direct relation to the fixed point discussed in Section~\ref{sec:a}.
In the following subsection we show that the spectrum of eigenoperators of
the two fixed points are identical, indicating that they describe
the same universal physics.

\subsection{Perturbations to finite  density}
\label{sec:perturb}

Having obtained the critical theory $\mathcal{S}_c$, we can now look
at all possible perturbations, and classify them according to their
renormalization group eigenvalues. It turns out that there are 3
relevant perturbations, and all other perturbations are irrelevant.
Anticipating the RG computation, we can arrange these perturbations
into eigenoperators of the RG transformation as
\begin{eqnarray}
\mathcal{S}_p & = & \int \dd \tau \dd^d x \Biggl\{ - \mu (\psi^\dagger_1
\psi_1 + \psi^\dagger_2 \psi_2 + 2 \Phi^\dagger \Phi) \nonumber \\
&& ~~~~~~~~~~~~~ + \delta  \Phi^\dagger \Phi - h (\psi^\dagger_1 \psi_1 -
\psi^\dagger_2 \psi_2 ) \Biggr\}
\end{eqnarray}
We can interpret the three relevant perturbations as the chemical
potential, $\mu$, a measure of the detuning away from unitary
scattering, $\delta$, and the `magnetic' field, $h$, which breaks
the symmetry between $\psi_1$ and $\psi_2$. It is a simple matter to
relate $\delta$ to the detuning $\nu$ in Eq.~(\ref{fr1}). The $T$
matrix for scattering between $\psi_1$ and $\psi_2$ is equal to
$g_0^2 G_\Phi$. We have to evaluate this on-shell, for incoming
particles with momenta $k_1$ and $k_2$, which then requires $G_\phi
(k=k_1+k_2, \omega = -i (k_1^2 /(2 m_1) + k_2^2 /(2 m_2)))$ in the
presence of a non-zero $\delta$. Using Eq.~(\ref{gp}) in $d=3$, and
comparing the result with Eq.~(\ref{fr1}), we conclude
\begin{equation}
\nu = \frac{2 \pi (m_1 + m_2)}{m_1 m_2} \frac{\delta}{g_0^2}
\end{equation}

The RG flow equations for the $\mu$, $\nu$, $h$ perturbations are
computed using the standard operator insertion methods. From this we
find the exact flow equations
\begin{eqnarray}
\frac{d\mu}{d\ell} &=& 2 \mu \nonumber \\
\frac{d\nu}{d\ell} &=& (2-\eta(g) ) \nu \nonumber \\
\frac{dh}{d\ell} &=& 2 h
\end{eqnarray}
Using the fixed point value of $\eta$ in Eq.~(\ref{exp}),
we see that the fixed point values of these scaling dimensions are
identical to those of the three relevant perturbations obtained in
Section~\ref{sec:a} for a field theory of atoms alone.

Universal correlations of the finite density quantum liquid can now
be computed as an expansion in the non-linear coupling $g=g^\ast$.
Because $g^{\ast 2 }$ is of order $(4-d)$, this leads to an
expansion in powers of $(4-d)$. Indeed results from such an
expansion have already been presented in Ref.~\onlinecite{son}.

As we noted in Section~\ref{sec:intro}, the resulting $(4-d)$
expansion does have some limitations. All fluctuation corrections to
the bare action $\mathcal{S}_c+\mathcal{S}_p$ are of order $(4-d)$,
and are consequently unable to change the mean-field ground state
obtained when $\mu$, $\nu$, and $h$ are of order unity. This has the
consequence that a direct expansion in powers of $(4-d)$ is only
able to describe the fully gapped superfluid, or the normal ground
state with a single Fermi surface. States with 2 Fermi surfaces, or
normal states at $T>0$ are only accessed after a redefinition of
parameter scales\cite{son,son2}, and are more conveniently accessed
by the method introduced in the following section.

\section{$1/N$ expansion for the $2N$ component Fermi gas}
\label{sec:largeN}

Following, Ref.~\onlinecite{wang}, we consider fermions
$\psi_\alpha$ with $\alpha = 1 \ldots 2N$ which have an Sp(2$N$)
invariant interaction with the boson $\Phi$. The critical theory for
this case is (for simplicity we take the fermions to have equal mass
$m$):
\begin{eqnarray}
\mathcal{S}_c = && \int \dd \tau \dd^d x \Biggl\{ \psi_\sigma^\dagger
\left( \frac{\partial}{\partial \tau} - \frac{\nabla^2}{2 m}
\right) \psi_\sigma \nonumber \\
&&~~~~~~~~+ \Phi^\dagger \left( \frac{\partial}{\partial \tau} -
\frac{\nabla^2}{4m } + r_{0c} \right) \Phi \nonumber \\
&~&~~~~~~~~- \frac{g_0}{2} \left( \Phi^\dagger \mathcal{J}^{\alpha
\beta} \psi_\alpha \psi_\beta + \mbox{H.c.} \right) \Biggr\}
\end{eqnarray}
where $\mathcal{J}^{\alpha\beta}$ is the Sp($2N$) invariant tensor
\begin{equation}
\mathcal{J} = \left( \begin{array}{cccccc}
  & 1 & & & & \\
-1 &  & & & & \\
 & &  & 1  & & \\
 & & -1 &  & & \\
 & &  & & \ddots & \\
 & & & & & \ddots
\end{array} \right)
\end{equation}
which is the generalization of the antisymmetric
$(4-d)^{\alpha\beta}$ tensor for SU(2) $\cong$ Sp(2). It is easy to
work out the exact $\beta$ function for this case, and we find
\begin{equation}
\frac{dg}{d\ell} =  \frac{(4-d)}{2} g - N g^3
\end{equation}
Note that the fixed point value of $g=g^\ast$ is of order $1/N$.
This implies that a $1/N$ expansion is possible at any value of
$(4-d)$.

Formally, the $1/N$ expansion is obtained by dropping the bare
quadratic terms in the critical theory for $\Phi$. This is because,
as we have seen in Eq,~(\ref{gp}), the self energy contributions to
the $\Phi$ Green's functions are more singular than the bare terms,
and dominate in the universal scaling limit.

It is now possible to explicitly write down the theory which
generates the universal properties to all orders in $1/N$.
Henceforth, all momentum integrals are implicitly evaluated in
dimensional regularization, and so we set $r_{0c}=0$ at the outset.
We use the action $\mathcal{S}_c + \mathcal{S}_p$ in its Sp($2N$)
invariant form, rescale $\Phi \rightarrow \Phi/g_0$, and finally
integrate the $\psi_\alpha$ fermions out completely. This yields a
theory for the single complex field $\Phi$ alone, with the effective
action
\begin{eqnarray}
\mathcal{S}_\Phi &=& N \Biggl\{ \frac{\delta}{g_0^2} \int \dd^d x
\dd\tau
 |\Phi (x, \tau) |^2 \nonumber \\
 &-&  \mbox{Tr} \ln \Bigl[
 \frac{\partial}{\partial \tau} - h + \left(- \frac{\nabla^2}{2 m} - \mu \right) \tau^3
 \nonumber \\ &~&~~~~~~~+
 \mbox{Re}[\Phi] \tau^1 + \mbox{Im}[\Phi] \tau^2
\Bigr] \Biggr\} \label{sPhi}
\end{eqnarray}
Here we have rescaled $\delta \rightarrow N\delta$, and
$\tau^{1,2,3}$ are the $2 \times 2$ matrices in Nambu space. With
the explicit factor of $N$ in front of the effective action for
$\Phi$, a $1/N$ expansion can be generated for all physical
properties.

The above approach may not appear very different from the usual
fluctuations-about-BCS theory of the paired Fermi gas. However, our
formal key observation, which follows from the existence of the RG
fixed point described in Section~\ref{sec:a} and~\ref{sec:am}, is
that upon evaluation of all integrals under dimensional
regularization, we automatically obtain the scaling functions
associated with crossovers across the Feshbach resonance.

The effective action (\ref{sPhi}) leads immediately to an explicit
$N=\infty$ result for the grand potential which obeys the universal
scaling form in Eq.~(\ref{scalef}). In principle, this result can be
obtained for an arbitrary spatial dependence of $\Phi$, which would
then allow for the spatially modulated `FFLO' phases (magnetized superfluids with Cooper pairs at a finite momentum $k_{F\uparrow}-k_{F\downarrow}$ determined by the Zeeman-split Fermi wave-vectors $k_{F\uparrow}$ and $k_{F\downarrow}$)\cite{fflo}; this extension is discussed briefly in Appendix~\ref{app:fflo}. We restrict our attention here to a constant $\Phi$, in which case
\begin{eqnarray}
&& \frac{\mathcal{F}^{(0)}}{N}  = \frac{m\nu}{4 \pi} |\Phi|^2 - \int
\frac{\dd^3 p}{(2 \pi)^3}
 \Biggl[ \nonumber \\
&&~~ T\ln \left(1 + e^{-(\sqrt{(p^2/(2m) - \mu)^2 + |\Phi|^2} -h)/T}
 \right) \label{finfty0} \\
 &&+ T \ln \left(1 + e^{-(\sqrt{(p^2/(2m) - \mu)^2 + |\Phi|^2} + h)/T}
 \right)  \nonumber \\
&&+ \sqrt{\left(\frac{p^2}{2m} - \mu\right)^2 + |\Phi|^2} -
    \left(\frac{p^2}{2m} - \mu\right)  \Biggr] \nonumber .
\end{eqnarray}
The superscript on $\mathcal{F}$ denotes that expression above is
divergent in the ultraviolet. As discussed above, the momentum
integral has to be evaluated in dimensional regularization, by a
process of analytic continuation from a convergent result for $d<2$
to $d>2$. We do this by adding and subtracting $m |\Phi |^2 /(p^2 +
\Lambda^2)$ from the integrand, where $\Lambda$ is an arbitrary
momentum scale. Then, the subtracting integrand is explicitly
convergent for $d=3$, while the compensating term can be evaluated
analytically by dimensional regularization. This yields
\begin{eqnarray}
&& \frac{\mathcal{F}}{N}  = \left(\frac{m\nu}{4 \pi} - \frac{m
\Lambda \Gamma(1-d/2)}{(4 \pi)^{d/2}} \right) |\Phi|^2 - \int
\frac{\dd^3 p}{(2 \pi)^3}
 \Biggl[ \nonumber \\
&&~~ T\ln \left(1 + e^{-(\sqrt{(p^2/(2m) - \mu)^2 + |\Phi|^2} -h)/T}
 \right) \label{finfty} \\
 &&+ T \ln \left(1 + e^{-(\sqrt{(p^2/(2m) - \mu)^2 + |\Phi|^2} + h)/T}
 \right)   \nonumber \\
&&+ \sqrt{\left(\frac{p^2}{2m} - \mu\right)^2 + |\Phi|^2} -
    \left(\frac{p^2}{2m} - \mu\right) -
\frac{m |\Phi|^2}{p^2+\Lambda^2} \Biggr] \nonumber .
\end{eqnarray}
where the dimensionally regularized term can be evaluated directly
in $d=3$. It can now be checked that the above expression is
independent of $\Lambda$, and is most easily evaluated at
$\Lambda=0$. The procedure above also shows that the subtraction is
only needed for the $|\Phi|^2$ term, but not for any of the higher
powers in $\Phi$.

The phase diagram is determined by minimizing Eq.~(\ref{finfty})
with respect to $\Phi$. The resulting free energy obeys
Eq.~(\ref{scalef}), and so yields a universal phase diagram as a
function of $\mu$, $h$, $\nu$ and $T$. We will present results for
aspects of this phase diagram, along with $1/N$ corrections, in the
subsections below.

\subsection{Superconductor to normal transition with increasing $T$}

At sufficiently high $T$, it is always the case that $\langle \Phi
\rangle = 0$. In this case, we can expand $\mathcal{S}_\Phi$ about
$\Phi =0$ for an arbitrary spatial and temporal dependence of
$\Phi$. To first order in $1/N$, we need to expand the action for
$\Phi$ to fourth order:
\begin{widetext}
\begin{eqnarray}
\frac{\mathcal{S}_\Phi}{N} &=& T \sum_{\omega_n} \int \frac{\dd^d
k}{(2 \pi)^d} K_2 (k, \omega_n) \Phi^\dagger (k, \omega_n) \Phi (k,
\omega_n ) \nonumber \\  &+& \frac{1}{2} \prod_{i=1}^3 \left( T
\sum_{\omega_{in}} \int \frac{\dd^d k_i}{(2 \pi)^d}\right)
\widetilde{K}_4 (k_i, \omega_{in}) \Phi^\dagger (k_1, \omega_{1n})
\Phi^\dagger (k_2, \omega_{2n}) \Phi(k_3, \omega_{3n}) \nonumber
\\&~&~~~~~~~~~~~~~~~~~~~~~~~~~~\times \Phi(k_1+k_2-k_3, \omega_{1n} + \omega_{2n} - \omega_{3n})
\end{eqnarray}
Here
\begin{equation}\label{K2}
K_2 (k, \omega_n ) = \frac{m\nu}{4 \pi} - \int \frac{\dd^d p}{(2
\pi)^d} \left(\frac{1 - f[(p+k)^2/(2m) -\mu -h] - f[p^2/(2m) - \mu +
h]}{-i \omega_n + p^2/(2m) + (p+k)^2 /(2m) - 2\mu} -
\frac{m}{p^2}\right) \,
\end{equation}
where $f(\epsilon) = 1/(e^{\epsilon/T} + 1)$ is the Fermi function,
and we have performed the dimensional regularization with the
subtraction of the $m/p^2$ term. Also, we will only need
$\widetilde{K}_4$ when two of its arguments have zero momenta and
frequency, in which case it equals $K_4 (k, \omega_n)$, given by
\begin{eqnarray}
K_4 (k, \omega_n) &=& \frac{1}{2} \int \frac{\dd^d p}{(2 \pi)^d}
\biggl[ I\left( (k-p)^2 / (2m) - \mu -h, p^2 /(2m) - \mu + h,
\omega_n \right) \nonumber \\&~&~~~~~~~~~~~~~+ I\left( (k-p)^2 /
(2m) - \mu + h, p^2 /(2m) - \mu - h, \omega_n \right) \biggr]
\end{eqnarray}
where
\begin{eqnarray}
I( \epsilon_1, \epsilon_2, \omega_n) &=&
\frac{\tanh(\epsilon_1/(2T)) + \tanh(\epsilon_2/(2T))}{ 4 \epsilon_2
( \epsilon_1 + \epsilon_2 - i \omega_n)^2} \nonumber \\ &+& \frac{
\tanh(\epsilon_1/(2T)) + \tanh(\epsilon_2/(2T)) - (\epsilon_2 /T)
{\rm sech}^2 (\epsilon_2 /(2T))}{ 8 \epsilon_2^2 ( \epsilon_1 +
\epsilon_2 - i \omega_n)} \nonumber \\ &-&
\frac{\tanh(\epsilon_1/(2T)) - \tanh(\epsilon_2/(2T))}{ 8
\epsilon_2^2 ( \epsilon_1 - \epsilon_2 - i \omega_n)}
\end{eqnarray}
\end{widetext}

Now we lower the temperature to critical temperature $T=T_c$ at
which there is an onset of superconductivity. To order $1/N$, the
condition for $T_c$ is given by
\begin{equation}
K_2 (0,0) + \frac{2 T}{N} \sum_{\omega_n} \int \frac{\dd^d k}{(2
\pi)^d} \frac{K_4 (k, \omega_n)}{K_2 (k, \omega_n)} = 0 \label{mass}
\end{equation}

We will restrict our analysis of the consequences of
Eq.~(\ref{mass}) to the balanced case at unitarity, $h=0$, $\nu =0$,
for which precise Monte Carlo results are also available. This will
allow us to test the accuracy of the $1/N$ expansion.

We tabulate the results of some integrals needed for the $1/N$
computation. At $N=\infty$, the transition is obtained by the
condition
\begin{equation}
K_2 (0,0)=0~~~~\Rightarrow~~~~\left. \frac{\mu}{T} \right|_{T=T_c} =
1.504476695 \label{tcinf}
\end{equation}
All numerical values tabulated below are for the value of $\mu$ in
Eq.~(\ref{tcinf}) and for $h=0$, $\nu=0$.
\begin{eqnarray}
- \frac{dK_2 (0,0)}{d\mu} &=& 0.018671 (2m)^{3/2} T^{-1/2} \nonumber \\
T \sum_{\omega_n} \int \frac{\dd^3 k}{8 \pi^3} \frac{K_4 (k, \omega_n)
}{K_2 (k, \omega_n)} &=& 0.0263 (2m)^{3/2} T^{1/2} \nonumber \\
2 \int \frac{\dd^3 k}{8 \pi^3} f\left(\frac{k^2}{2m}-\mu\right) &=&
0.096549 (2mT)^{3/2} \nonumber \\
2 \frac{d}{d\mu} \int \frac{\dd^3 k}{8 \pi^3}
f\left(\frac{k^2}{2m}-\mu\right) &=&
0.056179 (2m)^{3/2} T^{1/2} \nonumber \\
 -T \sum_{\omega_n} \int \frac{\dd^3 k}{8 \pi^3} \frac{d }{d\mu} \ln \left[K_2 (k, \omega_n) \right]
 &=& 0.2258 (2mT)^{3/2} \nonumber \\
2 T \int \frac{\dd^3 k}{8 \pi^3} \ln (1 + e^{-(k^2/2m-\mu)/T}) &=&
0.13188  (2m)^{3/2} T^{5/2} \nonumber \\
-T \sum_{\omega_n} \int \frac{\dd^3 k}{8 \pi^3}  \ln \left[K_2 (k,
\omega_n) \right]
 &=& 0.1357 (2m)^{3/2} T^{5/2} \nonumber \\
\label{numbers}
\end{eqnarray}

We make a few remarks about the numerical techniques used to obtain
the results in Eq.~(\ref{numbers}). All frequency summations were
evaluated on the imaginary frequencies. For all of the expressions,
the leading terms at large $\omega_n$, such as terms of order $\ln
(\omega_n)$, $1/\omega_n$, $1/\omega_n^{3/2}$, were obtained
explicitly. The summation over these leading terms was performed
analytically, along with the required time-splitting convergence
factors required for canonical bosons. After subtraction of these
leading terms, the summation over the remaining subleading terms was
performed numerically, and converged rapidly.

Using the first two results in Eq.~(\ref{numbers}) and the condition
Eq.~(\ref{mass}) we obtain the critical chemical potential:
\begin{equation}
\left. \frac{\mu}{T} \right|_{T=T_c} = 1.50448 + \frac{2.785}{N} +
\mathcal{O} (1/N^2). \label{chemical}
\end{equation}

The total density of fermions, $\rho$, is given to order $1/N$ by
\begin{equation}
\frac{\rho}{N} =  \int \frac{\dd^3 k}{8 \pi^3} \left[ 2
f\left(\frac{k^2}{2m}-\mu\right) - \frac{T}{N} \sum_{\omega_n}
\frac{d }{d\mu} \ln \left[K_2 (k, \omega_n) \right] \right]
\label{defrho}
\end{equation}
As is conventional, we relate  $\rho$ to a ``Fermi energy'',
$\varepsilon_F$, by
\begin{equation}
\varepsilon_F = \frac{(3 \pi^2 \rho/N)^{2/3}}{2m} \label{defef}
\end{equation}
Using the third, fourth, and fifth equations in Eq.~(\ref{numbers}),
the value of $\mu$ in Eq.~(\ref{chemical}) and the relations
Eq.~(\ref{defrho}) and (\ref{defef}) we obtain
\begin{equation}
 \left. \frac{\varepsilon_F}{T} \right|_{T=T_c}
= 2.01424 + \frac{5.317}{N} + \mathcal{O} (1/N^2) \label{ef}
\end{equation}

Finally, the pressure $P$ is the negative of the grand potential,
and so at order $1/N$
\begin{eqnarray}\label{Pressure}
\frac{P}{N} &=&  T\int \frac{\dd^3 k}{8 \pi^3} \Biggl[ 2 \ln (1 +
e^{-(k^2/2m-\mu)/T}) \nonumber \\
&~&~~~~~~~~~~~~- \frac{T}{N} \sum_{\omega_n} \ln \left[K_2 (k,
\omega_n) \right] \Biggr]. \label{defpress}
\end{eqnarray}
Using the third and the last two relations in Eq.~(\ref{numbers}),
the value of $\mu$ in Eq.~(\ref{chemical}), and
Eq.~(\ref{defpress}), we obtain
\begin{equation}
\left. \frac{P/N}{(2m)^{3/2} T^{5/2}} \right|_{T=T_c} = 0.13188 +
\frac{0.4046}{N} + \mathcal{O} (1/N^2) \label{press}
\end{equation}

Although the $1/N$ corrections in Eqs.~(\ref{chemical}), (\ref{ef})
and (\ref{press}), are quite large, direct evaluation of the
expressions at $N=1$ yields values that are in reasonable agreement
with recent Monte Carlo results \cite{troyer}, which obtained
$\mu/T_c = 3.247$, $\varepsilon_F /T_c = 6.579$, and $P/(2m^{3/2}
T_c^{5/2}) = 0.776$. The leading $N=\infty$ contributions in the
expansion are the contribution of fermion loops, while the $1/N$
corrections come from the boson loops: the distinct physical origins
of these contributions indicates that it is not meaningful to
compare their values as a test of the accuracy of the expansion. All
subsequent $1/N$  contributions are corrections to either the boson
or fermion loops, and the agreement with the Monte Carlo results is
an encouraging signal that such higher order corrections are indeed
small.

\subsubsection{Relationship to other work}

The structure of our leading $1/N$ corrections bears some similarity
to the recent work of Haussmann {\em et al\/} \cite{zwerger}. They
consider the single fermion loop (our $N=\infty$ terms) and the
single boson loop (our $1/N$ corrections) at an equal footing. Thus
their approach can be viewed as effectively similar to a
self-consistent $1/N$ computation. However, their approach breaks
down near $T_c$, and leads to unphysical results. Near $T_c$, the
value of $K_2 (0,0)$ approaches 0, and is also allowed to become
negative in a self-consistent approach. Indeed, a direct attempt to
obtain $T_c$ by solving Eq.~(\ref{mass}) self-consistently for
arbitrary $N$ leads immediately to this difficulty: we obtain $K_2
(0,0) < 0$ already for $T>T_c$, and the fluctuation propagator in
the second term has a negative `mass' and so is obtained by
performing a functional integral with an unphysical inverted
Gaussian weight. However, organizing the perturbation theory
order-by-order in $1/N$ (as we have done) avoids this difficulty,
and leads to a well-defined and controlled expansion for the
thermodynamic properties at $T_c$, with all boson loop propagators
always obeying $\mbox{Re}[K_2 (k,\omega_n)] \geq 0$.

\subsection{Phase diagram at zero temperature}

We will now present some results on the universal phase diagram at
$T=0$. This is determined by 3 parameters: $\mu$, $\nu$, and $h$.
However, these parameters are dimensionful, and the phase diagram
can only depend only upon {\em two\/} independent ratios of these
parameters which have zero scaling and engineering dimensions. From
Eq.~(\ref{scalef}), we can deduce that the phase diagram depends
upon the ratios of $\mu$, $h$, and $\nu^2 /(2m)$. The sign of $h$ is
immaterial, while those of $\mu$ and $\nu$ are certainly
significant, and one or more of these parameters may be zero;
consequently we cannot explore the complete phase diagram in a
single two-dimensional plot.

First, let us begin with some general considerations at $h=0$. As
long as the detuning, $\nu$, is positive, it will pay to have
particles in the ground state of the grand canonical ensemble only
for $\mu > 0$. However, for $\nu < 0$, the negative detuning implies
there is a molecular bound state\cite{rembert} at energy $-\nu^2/m$.
So we can expect a non-zero density of particles for $2\mu
> - \nu^2 / m$. This reasoning led to the phase diagram shown
in Fig.~\ref{zeroh}. We will see shortly that this threshold value
of $\mu$ is indeed obtained in the $N=\infty$ theory.

For $h \neq 0$, we will limit our analysis here to phases with
$\langle \Phi \rangle$ spatially uniform in the $N = \infty$ theory.
At $N=\infty$ and $T=0$, the free energy density in
Eq.~(\ref{finfty}) at $\Lambda=0$ reduces to:
\begin{eqnarray}\label{F00}
\frac{\mathcal{F}}{N} & = & \frac{m\nu}{4\pi}|\Phi|^2 -
  \int \frac{\dd^3 p}{(2\pi)^3} \Biggl\lbrack
  \left( h - \sqrt{\left( \frac{p^2}{2m} - \mu \right)^2 + |\Phi|^2} \right)
    \nonumber \\ & & \times
  \theta \left( h - \sqrt{ \left( \frac{p^2}{2m} - \mu \right)^2 + |\Phi|^2} \right)
    \\
& & + \sqrt{\left( \frac{p^2}{2m} - \mu \right)^2 + |\Phi|^2} -
  \left( \frac{p^2}{2m} - \mu \right) - \frac{m|\Phi|^2}{p^2} \nonumber
  \Biggr\rbrack \ ,
\end{eqnarray}
where $h>0$ was assumed without loss of generality. At $h=0$, we can
check for the transition in Fig.~\ref{zeroh} by expanding
Eq.~(\ref{F00}) to order $| \Phi |^2$ for $\mu < 0$. This yields
\begin{equation}
\frac{\mathcal{F}}{N} = |\Phi |^2 \left[
 \frac{m\nu}{4 \pi} - \int \frac{\dd^3 p}{8 \pi^3} \left(
 \frac{1}{2(p^2/(2m) + |\mu|)} - \frac{m}{p^2} \right) \right] +
\ldots
\end{equation}
Evaluating the integral, we observe that the co-efficient of $|\Phi
|^2$ changes sign at $\mu = - \nu^2 /(2m)$, thus establishing the
location of the phase boundary in Fig.~\ref{zeroh} at $N=\infty$. It
is also easy to see that there are no corrections to the position of
this phase boundary to all orders in $1/N$: the $\Phi$ propagator
has spectral weight only at positive frequencies at the phase
boundary (including a pole at $\omega =k^2/(4m)$), and this implies
that all loop corrections vanish.

Let us now turn to $h \neq 0$. The superfluid phase in
Fig.~\ref{zeroh} has a gap, equal to $|\Phi|$ (`spin' excitations),
and consequently a non-zero $|h| \leq |\Phi|$ has no effect on the
superfluid phase. The density of the particles remains balanced in
this regime, and depends only upon the values of $\mu$ and $\nu$.

\begin{figure}
\includegraphics[width=3.1in]{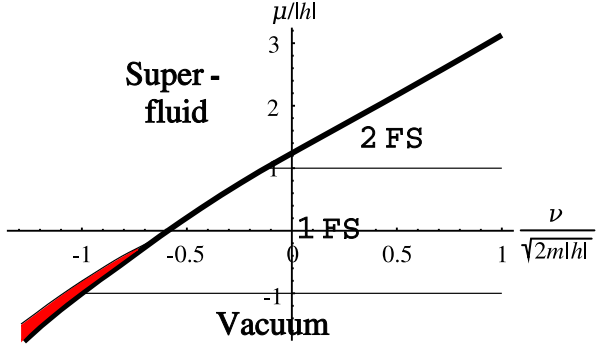}
\caption{\label{finiteh}(color online) Universal phase diagram at
$T=0$, $h \neq 0$, and $N = \infty$. The axes have been scaled by
$|h|$ to the dimensionless parameters $\mu/|h|$ and $\nu/\sqrt{2m
|h|}$. The density in the superfluid is balanced, except in the
shaded (red) region representing a `magnetized superfluid', which
also has a single (1 $\times N$) Fermi surface of (say) up spin
fermions (for $h>0$). The 1FS and 2FS phases are non-superfluid
states with $N$ and $2N$ Fermi surfaces respectively. The thick line
is a first order quantum phase transition, while the thin lines are
second order transitions.}
\end{figure}

We show a finite $h$ phase diagram in Fig.~\ref{finiteh}, obtained
by determining the global minimum of the $N=\infty$ free energy in
Eq.~(\ref{F00}) with respect to variations in a space-independent
$\Phi$ --- a remarkable amount of information emerges from the
minimization of this single function. The structure of this
universal phase diagram has similarities to that obtained in the
resonance-width expansion\cite{leo}.
Because the physics is invariant under the change of sign of $h$, we
plot the phases as a function of the dimensionless parameters
$\mu/|h|$ and $\nu/\sqrt{2 m |h|}$ In addition to the conventional
gapped superfluid state with $|h| < |\Phi|$ and balanced densities,
there is also  a more exotic superfluid phase in the phase diagram.
The Luttinger theorems described in Ref.~\onlinecite{kun2}, require
that any ground state with an unbalanced density must have at least
one Fermi surface (as long as the `spin' rotation symmetry about the
`field' $h$, pointing in the $z$ direction in spin space, remains
unbroken). The shaded region in Fig~\ref{zeroh} is such an
unbalanced superfluid phase, which has a single Fermi surface of one
of the fermion species (that is, $N$ rather than $2N$) co-existing
with the superfluid condensate. Modulated FFLO phases\cite{fflo}
have $\langle \Phi \rangle \neq 0$ and space-dependent, and have
Fermi surfaces: we briefly discuss such states in
Appendix~\ref{app:fflo}, but have not yet undertaken the numerical
analysis to determine their structure and stability---we will do so
in future work.

\begin{figure}
\includegraphics[width=3.1in]{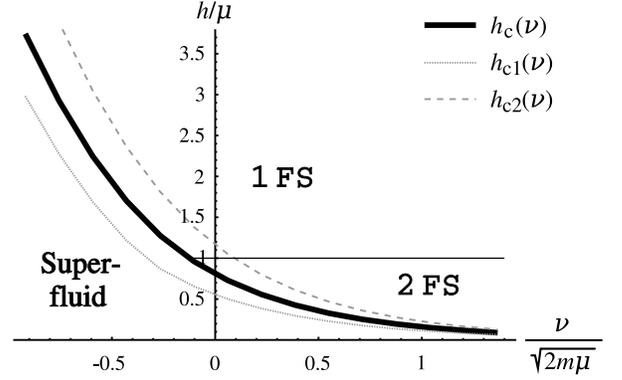}
\caption{\label{PhDiag0} The same universal, zero temperature phase
diagram as in Fig.~\ref{finiteh}, but for $\mu > 0$. Now we have
scaled the axes by $\mu$, and they measure detuning and the field
respectively. The first order phase transition between the
superfluid and normal phases occurs at $h=h_c(\nu)$, plotted as the
thick line; $h_c(0)=0.807125\mu$. The dashed faint line
denotes the ``upper critical field'' $h_{c2}(\nu)$ (equal to the
fermion BCS pairing gap), below which the superfluid may be found at
least as a metastable state. Similarly, the dotted faint line
denotes the ``lower critical field'' $h_{c1}(\nu)$, above which the
normal state may be found at least as a metastable state.}
\end{figure}

\begin{figure}
\includegraphics[width=3.1in]{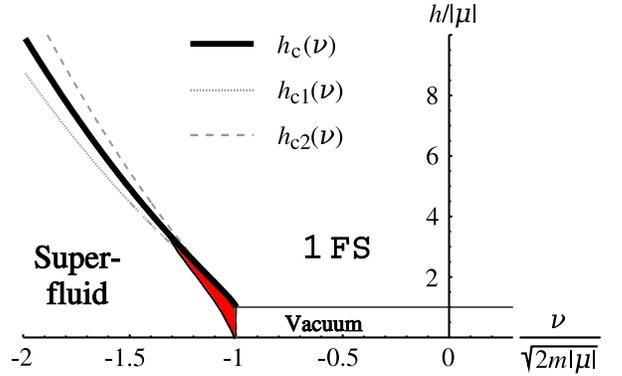}
\caption{\label{PhDiag0b}(color online) As in Fig.~\ref{PhDiag0}, but for $\mu <
0$. The shaded (red) region is a `magnetized superfluid'.}
\end{figure}

Other views of the same $N=\infty$ phase diagram are presented in
Figs.~\ref{PhDiag0} and~\ref{PhDiag0b}
The only other stable states we find in our $N=\infty$ theory (with
$\Phi$ space-independent) are the normal states with $\langle \Phi
\rangle = 0$. This appears across a first-order transition indicated
by the thick line in Fig.~\ref{PhDiag0}. As usual, the existence of
this first-order transition as a function of chemical potential
means that a system with fixed total density will undergo phase
separation. In the present $N=\infty$ theory, one of the phase
separated phases will be a superfluid with no population imbalance
(fluctuations are not included at $N\to\infty$). The other phase
will be one of the normal states in Fig.~\ref{PhDiag0}, which are
present for $h>h_c(\nu)$. This normal state has either $N$ or $2N$
Fermi surfaces, depending on whether the external field is strong
enough to fully polarize the fermions. The critical field at the
Feshbach resonance is $h_c(0) = 0.807125\mu$, and in the range
$h_c(0)<h<\mu$ the normal state has $2N$ Fermi surfaces.

It is also interesting to note that the local minimum of free energy
at $\Phi \neq 0$ survives at fields $h$ as high as $h_{c2}(\nu)$
shown in Fig.~\ref{PhDiag0} as the dashed line. This local minimum
of free energy is global only for $h < h_c(\nu) < h_{c2}(\nu)$.
Similarly, the free energy has a local minimum at $\Phi=0$ for $h > h_{c1} (\nu)$.
Therefore, metastable superfluid and normal phases can exist in the range $h_{c1}(\nu) < h < h_{c2}(\nu)$.

We now consider $1/N$ corrections to the equation of state of the
normal regions of the phase diagram. The pressure is calculated from
Eq.~(\ref{Pressure}), which at zero temperature takes the form:
\begin{eqnarray}
\frac{P}{N} & = & \frac{\mu (2m\mu)^{\frac{3}{2}}}{15\pi^2}
\Biggl\lbrack
  \theta\left(1-\frac{h}{\mu}\right) \left(1-\frac{h}{\mu}\right)^{\frac{5}{2}} \label{pn} \\
& + &  \theta\left(1+\frac{h}{\mu}\right)
\left(1+\frac{h}{\mu}\right)^{\frac{5}{2}} \Biggr\rbrack +
\frac{\delta P}{N} \ . \nonumber
\end{eqnarray}
The calculation of the $1/N$ correction $\delta P$ is presented in
the Appendix~\ref{appK2}. These $1/N$ corrections are non-zero at
$T=0$ only in the region where both fermion species are present in
the ground state {\em i.e.\/} in the 2FS state. This state appears
only for $\mu > 0$, and so we assume a positive $\mu$ below. The
scaling form in Eq.~(\ref{scalef}) implies that at $T=0$ this $1/N$
correction can be written in $d=3$ in the form ($\mu > 0$)
\begin{equation}
\frac{\delta P}{N} = \mu (2m\mu)^{3/2} F_{\delta P} \left(
\frac{h}{\mu}, \frac{\nu}{\sqrt{2m\mu}} \right)
\end{equation}
where $F_{\delta P}$ is a universal function. Our results for the
function $F_{\delta P}$ are derived in Appendix~\ref{appK2}, and
plotted in Fig.~\ref{PressurePlot} as a function of $h/\mu$ at
various values of detuning from the Feshbach resonance.
\begin{figure}
\includegraphics[width=3in]{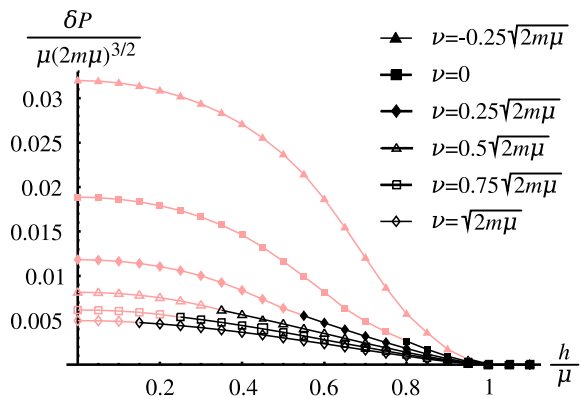}
\caption{\label{PressurePlot}(color online) $1/N$ correction to
pressure in the normal state ($h>h_{c1}(\nu)$) at zero temperature
for $\mu > 0$. In the limit $N\to\infty$ the faded pink portions of
the curves should be discarded since they lie within the superfluid
phase.}
\end{figure}
Virtual pairing fluctuations tend to increase pressure, and more so
for longer lived pairs. However, for $h>\mu$ the fermion gas is
fully polarized and no pairing fluctuations remain to contribute the
increase of pressure. Note that the shown $\delta P$ is valid only
when the superfluidity is destroyed. This happens for $h>h_c(\nu)$,
which in the case of infinite $N$ corresponds to the dark portions
of the curves in the Fig.~\ref{PressurePlot}.

The results in Eq.~(\ref{pn}) and Fig~\ref{PressurePlot} can have a
number of important consequences. Among the most important is that
the position of the phase boundaries in Fig.~\ref{PhDiag0} will be
moved significantly, especially for negative detuning; additional
phases not present at $N=\infty$ could also appear. However,
determining this change requires determination of the $1/N$
correction to the energy of the superfluid state: this is a
computation similar to that in Appendix~\ref{appK2}, but more
involved because of the broken symmetry in the superfluid state.
This computation we also defer to future work, as we discuss further
in Section~\ref{sec:conc}. Also, the equation of state of the normal
state implied by Fig~\ref{PressurePlot} has significant consequences
for recent experiments\cite{martin,randy}, and these will be
described in a separate publication.

\section{Conclusions}
\label{sec:conc}

This paper has presented a unified understanding of quantum liquids
near unitarity in various dimensions using a renormalization group
approach. The universal behavior is associated with RG fixed points
describing the interactions of a few `atoms' and `molecules'. This
fixed point of the zero density theory is\cite{book} nevertheless a
useful and powerful starting point for understanding properties of
the quantum liquid with a finite density of particles. We view the
RG fixed point as a description of a quantum critical point obtained
as the chemical potential of the particles is varied. The quantum
(multi-)critical point then describes a quantum phase transition (in
the grand canonical ensemble) between a vacuum ground state with
zero density of particles, and a finite density quantum liquid; see
Fig~\ref{zeroh}. We have shown how this perspective allows us to
bring the standard and well-developed techniques of critical
phenomena to the computation of the phase diagram of such liquids.

Our principal application of this method was to the attractive Fermi
gas in $d=3$. The fixed point theory can be described in either a
$(d-2)$ or a $(4-d)$ or a $1/N$ expansion (for models with Sp($2N$)
symmetry). We argued that the $1/N$ expansion was best suited for
the purposes of describing the onset of the non-superconducting
ground state with increasing temperature or density imbalance, and
presented results to order $1/N$. We also obtained $1/N$
contributions to the equation of state of the normal state.

A number of additional results can be obtained in the $1/N$
expansion, and these we hope to address in separate publications:
\begin{itemize}
\item
Single particle Green's functions can also be computed in the $1/N$
expansion by standard methods. In the normal state, these will yield
information on the consequences of pairing fluctuations.
\item
The existence and properties of the modulated FFLO phases are, in
principle, also described by universal scaling forms such as
Eq.~(\ref{scalef}). The form of the $N=\infty$ free energy of such
phases is discussed in Appendix~\ref{app:fflo}, and a numerical
analysis of the large parameter space of this result is needed.
\item
Apart from the existence of FFLO and other phases, the phase
boundaries in Fig~\ref{PhDiag0} will also be shifted by $1/N$
corrections, especially in the regime of negative detuning. To
compute these, we need the $1/N$ corrections to the free energy in
the uniform superfluid phase. This will have an expression which
generalizes Eq.~(\ref{deltaP}) to the broken symmetry phase, and can
be computed in a similar manner.
\item
The equations of state obtained herein, and as indicated above, are
expressed in terms of chemical potentials. These can be converted to
equations of state as a function of particle densities by standard
thermodynamic methods. The local density approximation can then be
used to convert the equation of state into particle density profiles
in the presence of an external trapping potential.
\end{itemize}

\acknowledgments S.S. would like to thank Kun Yang for useful
discussions and a previous collaboration on related
issues.\cite{kun2} We thank Leo Radzihovsky for valuable comments on
the manuscript, for pointing out that a preliminary phase diagram
was incomplete, and for informing us about parallel work by
M.~Veillette, D.~Sheehy, and him on the RG approach and the $1/N$
expansion\cite{rvs}; where they overlap, our results agree with
theirs. We also thank E.~Demler, M.~Randeria, D.~Son, and
M.~Zwierlein for useful discussions. This research was supported by
the NSF under grant DMR-0537077. We thank Aspen Center for Physics
for hospitality.

\appendix

\section{$N=\infty$ theory for FFLO states}
\label{app:fflo}

This appendix generalizes the $N=\infty$ result in
Eq.~(\ref{finfty}) to states with a spatially dependent $\Phi$.

We expect that the optimum solution for $\Phi(x)$ will have the
symmetry of some Bravais lattice, and so write
\begin{equation}
\Phi (x)  = \sum_{{\bf G}} \Phi_{\bf G} e^{i {\bf G} \cdot {\bf x}}
\end{equation}
where ${\bf G}$ are the reciprocal lattice vectors. First, we need
the energy eigenvalues of the Bogoliubov-de Gennes equations in the
presence of such a $\Phi (x)$. We can expand the eigenmodes
$(u(x),v(x))$ also in plane-wave Bloch eigenstates
\begin{equation}
u(x) = e^{i {\bf p} \cdot {\bf x}} \sum_{\bf G} u_{\bf G} e^{i {\bf
G} \cdot {\bf x}},
\end{equation}
and similarly for $v(x)$. Then the eigenvalue equation is
\begin{eqnarray}
\left( \frac{({\bf p}+{\bf G})^2}{2m} - \mu \right) u_{\bf G} +
\sum_{{\bf G}'} \Phi_{{\bf G}-{\bf G}'}
v_{{\bf G}'} &=& \epsilon_{\bf p} (\Phi) u_{\bf G} \nonumber \\
-\left( \frac{({\bf p}+{\bf G})^2}{2m} - \mu \right) v_{\bf G} +
\sum_{{\bf G}'} \Phi^\ast_{{\bf G}-{\bf G}'} u_{{\bf G}'} &=&
\epsilon_{\bf p} (\Phi) v_{\bf G} \nonumber ,
\end{eqnarray}
where $\epsilon_{\bf p} (\Phi)$ are the Bloch eigenenergies. We now
have to choose a truncation of the set of reciprocal lattice
vectors, and then numerically diagonalize the equations above to
obtain these eigenenergies. An infinite set of positive
eigenenergies will be obtained for each ${\bf p}$ in the first
Brillouin zone. In an extended zone scheme, these eigenenergies can
be rearranged to obtain a single-valued and positive function of
${\bf p}$, $\epsilon_{\bf p} (\Phi)$, where ${\bf p}$ extends over
all real values in $d$-dimensional momentum space. This function can
be chosen so that for $\Phi$ constant, $\epsilon_{\bf p} (\Phi) =
((p^2/(2m)-\mu)^2 + |\Phi|^2)^{1/2}$.

Now, it is easy to see that the generalization of
Eq.~(\ref{finfty0}) is
\begin{eqnarray}
&& \frac{\mathcal{F}^{(0)}}{N}  = \frac{m\nu}{4 \pi} \sum_{\bf G}
|\Phi_{\bf G}|^2 - \int \frac{\dd^3 p}{(2 \pi)^3}
 \Biggl[ \nonumber \\
&&~~ T\ln \left(1 + e^{-(\epsilon_{\bf p} (\Phi) -h)/T}
 \right)  + T \ln \left(1 + e^{-(\epsilon_{\bf p} (\Phi) + h)/T}
 \right)  \nonumber \\
&&~~~~~~~~~+ \epsilon_{\bf p} (\Phi) -
    \left(\frac{p^2}{2m} - \mu\right)  \Biggr] \label{finfty0a} .
\end{eqnarray}
where the integral is over all momenta. As in Eq.~(\ref{finfty0})
this expression suffers from an ultraviolet divergence, and we cure
this by dimensional regularization. At large $p$, the values of
$\epsilon_{\bf p} (\Phi)$ can be obtained by perturbation theory in
$\Phi$, and the divergent pieces only involve terms to second order
in $\Phi$. Adding and subtracting this divergent term to
Eq.~(\ref{finfty0a}), and analytically performing the dimensional
regularization as in Eq.~(\ref{finfty}) at $\Lambda = 0$, we obtain
\begin{eqnarray}
&& \frac{\mathcal{F}}{N}  = \frac{m\nu}{4 \pi} \sum_{\bf G}
|\Phi_{\bf G}|^2 - \int \frac{\dd^3 p}{(2 \pi)^3}
 \Biggl[ \nonumber \\
&&~~ T\ln \left(1 + e^{-(\epsilon_{\bf p} (\Phi) -h)/T}
 \right)  + T \ln \left(1 + e^{-(\epsilon_{\bf p} (\Phi) + h)/T}
 \right)  \nonumber \\
&&~~~~~~~~~~+ \epsilon_{\bf p} (\Phi) -
    \left(\frac{p^2}{2m} - \mu\right)  - \frac{m}{p^2} \sum_{\bf G} |\Phi_{\bf G}|^2 \Biggr] \label{finftya} .
\end{eqnarray}
It now remains to numerically minimize Eq.~(\ref{finftya}) over the
set of values of $\Phi_{\bf G}$.

\section{Equation of state at $T=0$}\label{appK2}

Here we present some details regarding the $1/N$ corrections to the
equation of state in the absence of condensate at zero temperature.
These corrections appear only in the normal states with two ($2\times N$) Fermi surfaces, so that we assume $\mu>0$. The equation of state up to the order of $1/N$ is generally obtained from Eq.~(\ref{Pressure}) generalized to finite external fields $h$.

At $T=0$ and in the absence of condensate it is possible to
analytically calculate $K_2$. Consider first the imaginary part of
Eq.~(\ref{K2}) after analytic continuation to real frequencies
$i\omega \to \omega + i0^+$:
\begin{eqnarray}\label{ImK2}
&& -\frac{1}{\pi} \textrm{Im} \lbrace K_2(k,\omega) \rbrace =
  \int \frac{\dd^3 p}{(2\pi)^3}
  \Biggl\lbrack 1 - \theta\left(h-\frac{p^2}{2m}+\mu\right) \nonumber \\
&& ~~~~~ - \theta\left(-h-\frac{(\bf{p}+\bf{k})^2}{2m}+\mu\right) \Biggr\rbrack
  \times \\
&& ~~~~~ \delta\left( \omega +2\mu - \frac{p^2}{2m} - \frac{(\bf{p}+\bf{k})^2}{2m}\right) \ . \nonumber
\end{eqnarray}
Note that in the rest of this appendix we will denote by
$K_2(k,\omega)$ the dependence on \emph{real} frequency $\omega$,
which is a slight change of notation from Eq.~(\ref{K2}). Upon the
shift of variables $\bf{p} \to \bf{p} - \bf{k}/2$, the delta
function fixes the magnitude of $\bf{p}$:
\begin{equation}\label{pfn}
p = p(k,\omega) \equiv \sqrt{m(2\mu+\omega)-\frac{k^2}{4}} \ ,
\end{equation}
while the remaining integral over spatial direction of $\bf{p}$ is
carried out easily:
\begin{equation}\label{K2b}
\textrm{Im} \lbrace K_2(k,\omega) \rbrace = \textrm{Im} \lbrace
K_2'(k,\omega) \rbrace
  - \frac{mp}{4\pi} \theta\left(p^2\right)
\end{equation}
where
\begin{eqnarray}
&& \textrm{Im} \lbrace K_2'(k,\omega) \rbrace = \frac{mp}{8\pi}
   \theta\left(p^2\right) \\
&& ~~~~~~~~ \times \left\lbrack
   R\left(\frac{m(\omega+2h)}{pk}\right) +
   R\left(\frac{m(\omega-2h)}{pk}\right) \right\rbrack
   \nonumber \ ,
\end{eqnarray}
and
\begin{equation}
R(x) = \left\lbrace
  \begin{array}{lcl}
    2   & \quad , \quad & x \leq -1 \\
    1-x & \quad , \quad & -1 \leq x \leq 1 \\
    0   & \quad , \quad & x \geq 1
  \end{array} \right\rbrace \ .
\end{equation}
The real part of $K_2(k,\omega)$ can be obtained from the
requirement that $K_2$ be analytic in the upper complex half-plane
of $\omega$. Hence, the Kramers-Kronig relations will be useful, but
they can be applied only to the well behaved portion
$K_2'(k,\omega)$. Fortunately, it is easy to recognize that the
portion of $K_2(k,\omega)$ whose imaginary part diverges when
$\omega\to\infty$ (the second term in ~(\ref{K2b})) is just a square
root function:
\begin{eqnarray}\label{K2c}
&& \frac{m}{4\pi}\sqrt{\frac{k^2}{4}-m(2\mu+\omega+i0^+)} = \\
&& ~~~~~~~~~~~ \frac{m\sqrt{-p^2}}{4\pi} \theta\left(-p^2\right) -
          i \frac{mp}{4\pi} \theta\left(p^2\right) \ . \nonumber
\end{eqnarray}

\begin{figure}
\includegraphics[width=3.2in]{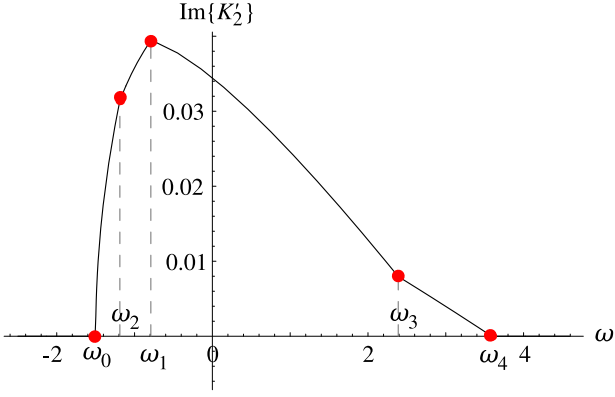}
\caption{\label{K2plot}(color online) A typical plot of
$\textrm{Im}\lbrace K_2'(k,\omega) \rbrace$ versus frequency (here
$\mu=1$, $m=0.5$, $h=0.2$, $k=1$). There are up to five kink
frequencies, listed in ~(\ref{kinks}), including the end-points.
$\textrm{Im}\lbrace K_2'(k,\omega) \rbrace$ is non-zero only in a
finite interval.}
\end{figure}

We now determine the real part of $K_2'(k,\omega)$ using the
Kramers-Kronig relation:
\begin{equation}\label{KramersKronig}
\textrm{Re}\lbrace K_2'(k,\omega)\rbrace = \frac{1}{\pi}
\int\limits_{-\infty}^{\infty}
  \dd\omega' \frac{\mathbb{P}}{\omega'-\omega}
  \textrm{Im}\lbrace K_2'(k,\omega')\rbrace \ .
\end{equation}
A typical plot of $\textrm{Im}\lbrace K_2'(k,\omega)\rbrace$ is
shown in Fig.~\ref{K2plot}. Depending on the values of $\mu$, $h$
and $k$ there are up to five frequencies where $\textrm{Im}\lbrace
K_2'(k,\omega)\rbrace$ has kinks:
\begin{eqnarray}\label{kinks}
\omega_0 & = & \frac{k^2}{4m}-2\mu \\
\omega_1 & = & \omega_0 + \frac{1}{m} \left( \frac{k}{2} -
\sqrt{2m(\mu+h)} \right)^2
  \nonumber \\
\omega_2 & = & \omega_0 + \frac{1}{m} \left( \frac{k}{2} -
\sqrt{2m(\mu-h)} \right)^2
  \nonumber \\
\omega_3 & = & \omega_0 + \frac{1}{m} \left( \frac{k}{2} +
\sqrt{2m(\mu-h)} \right)^2
  \nonumber \\
\omega_4 & = & \omega_0 + \frac{1}{m} \left( \frac{k}{2} +
\sqrt{2m(\mu+h)} \right)^2
  \nonumber \ .
\end{eqnarray}
We evaluate the integral ~(\ref{KramersKronig}) piecewise between
these kink frequencies. Knowing both the real and imaginary parts of
$K_2'(k,\omega)$ for real frequencies, together with ~(\ref{K2b})
and ~(\ref{K2c}), allows reconstruction of the full function
$K_2(k,\omega)$, analytic in the upper complex half-plane of
$\omega$:
\begin{widetext}
\begin{eqnarray}\label{K2full}
&& K_2(k,\omega) = \frac{m\nu}{4\pi} +
                   \frac{m}{4\pi}\sqrt{\frac{k^2}{4}-m(2\mu+\omega)} \\
&& ~~~~~~ + \frac{m}{4\pi^2}\theta(\mu+h) \Biggl\lbrace
     \theta \left( 2m(\mu+h)-\frac{k^2}{4} \right) \left\lbrack 2p(\omega_1)
   + p(\omega) \ln \left( -\frac{p(\omega_1)-p(\omega)}{p(\omega_1)+p(\omega)} \right)
     \right\rbrack \nonumber \\
&& ~ - \frac{p(\omega)}{2} \left\lbrack
      \ln \left( \frac{p(\omega_4)+p(\omega)}{p(\omega_4)-p(\omega)} \right) -
      \ln \left( \frac{p(\omega_1)+p(\omega)}{p(\omega_1)-p(\omega)} \right)
     \right\rbrack
    + p(\omega_4) - p(\omega_1) - \frac{m}{2k} \left\lbrack \omega_4 - \omega_1
      + (\omega-2h) \ln \left( \frac{\omega_4-\omega}{\omega_1-\omega} \right)
     \right\rbrack \Biggr\rbrace \nonumber \\
&& ~~~~~~ + \frac{m}{4\pi^2}\theta(\mu-h) \Biggl\lbrace
     \theta \left( 2m(\mu-h)-\frac{k^2}{4} \right) \left\lbrack 2p(\omega_2)
   + p(\omega) \ln \left( -\frac{p(\omega_2)-p(\omega)}{p(\omega_2)+p(\omega)} \right)
     \right\rbrack \nonumber \\
&& ~ - \frac{p(\omega)}{2} \left\lbrack
      \ln \left( \frac{p(\omega_3)+p(\omega)}{p(\omega_3)-p(\omega)} \right) -
      \ln \left( \frac{p(\omega_2)+p(\omega)}{p(\omega_2)-p(\omega)} \right)
     \right\rbrack
    + p(\omega_3) - p(\omega_2) - \frac{m}{2k} \left\lbrack \omega_3 - \omega_2
      + (\omega+2h) \ln \left( \frac{\omega_3-\omega}{\omega_2-\omega} \right)
     \right\rbrack \Biggr\rbrace \nonumber
\end{eqnarray}
\end{widetext}
Here, $p(\omega)$ is given by ~(\ref{pfn}), where dependence on $k$
is assumed and omitted for brevity.

\begin{figure}
\includegraphics[width=2in]{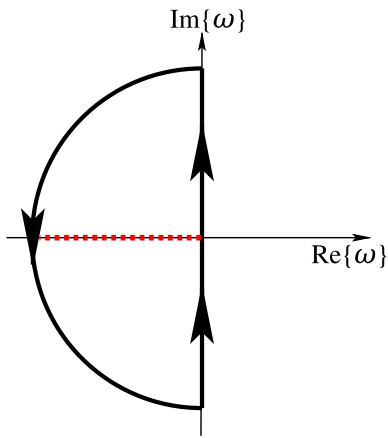}
\caption{\label{CPath}(color online) The complex path of $\omega$
along which $\ln K_2(k,\omega)$ is integrated out in ~(\ref{deltaP})
(the radius of the semi-circle is infinite). The dashed red line
shows the branch-cut of the logarithm.}
\end{figure}

Now we can calculate the $1/N$ correction to pressure in the normal
state at zero temperature. For this purpose we substitute
~(\ref{K2full}) into ~(\ref{Pressure}):
\begin{equation}\label{deltaP}
\frac{\delta P}{N} = - \frac{1}{N} \int \frac{\dd^3 k}{(2\pi)^3}
  \int\limits_{-\infty}^{\infty} \frac{\dd\omega}{2\pi}
  e^{i\omega 0^+} \ln \left\lbrack K_2(k,i\omega) \right\rbrack \ .
\end{equation}
The change of variables $i\omega \to \omega$ allows the frequency
integral to be evaluated on the closed path in complex plane shown
in Fig.~\ref{CPath}. Avoiding the branch-cut of the logarithm, and
exploiting the analyticity of $K_2(k,\omega)$, the integral above
reduces to:
\begin{eqnarray}
\frac{\delta P}{N} & = & \frac{i}{N} \int \frac{\dd^3
k}{(2\pi)^3}
  \int\limits_{-\infty}^{0} \frac{\dd\omega}{2\pi}
  \ln \left( \frac{K_2(k,\omega-i0^+)}{K_2(k,\omega+i0^+)} \right) \qquad\quad \\
& = & \frac{1}{N} \frac{1}{2\pi^3} \int\limits_0^{\infty} \dd k
  \int\limits_{-\infty}^{0} \dd \omega
  k^2 \arg \left\lbrack K_2(k,\omega+i0^+) \right\rbrack \ . \nonumber
\end{eqnarray}
In the second line we have used the fact that $K_2(k,\omega-i0^+) =
K_2^*(k,\omega+i0^+)$. This integral has to be evaluated
numerically. The plot of pressure correction at zero temperature is
shown in Fig.~\ref{PressurePlot}.

\end{document}